\documentclass[journal]{IEEEtran}
\usepackage{graphicx}
\usepackage{algorithm}
\usepackage{algpseudocode}
\usepackage[table,xcdraw]{xcolor}
\usepackage{multicol}
\usepackage{multirow}
\ifCLASSINFOpdf
\else
\fi
\usepackage{amsmath}
\usepackage{mathtools}
\usepackage{tikz}

\newcommand\copyrighttext{%
  \footnotesize This work has been submitted to the IEEE for possible publication. Copyright may be transferred without notice, after which this version may no longer be accessible}
\newcommand\copyrightnotice{%
\begin{tikzpicture}[remember picture,overlay]
\node[anchor=south,yshift=10pt] at (current page.south) {\fbox{\parbox{\dimexpr\textwidth-\fboxsep-\fboxrule\relax}{\copyrighttext}}};
\end{tikzpicture}%
}

\begin{document}
%
\title{Deep reinforcement learning for automatic run-time adaptation of UWB PHY radio settings}
%
%
%

\author{Dieter Coppens, Adnan Shahid, \IEEEmembership{Senior member , IEEE}, Eli De Poorter
\thanks{D. Coppens, A. Shahid, and E. De Poorter are with IDLab, Department of Information Technology, Ghent University—imec, 9052 Ghent,
Belgium (e-mail: dieter.coppens@ugent.be; eli.depoorter@ugent.be;
adnan.shahid@ugent.be}
}

%
%

\markboth{Journal of \LaTeX\ Class Files,~Vol.~14, No.~8, August~2015}%
{Shell \MakeLowercase{\textit{et al.}}: Bare Demo of IEEEtran.cls for IEEE Journals}
%



\maketitle
\copyrightnotice
\begin{abstract}
Ultra-wideband technology has become increasingly popular for indoor localization and location-based services. This has led recent advances to be focused on reducing the ranging errors, whilst research focusing on enabling more reliable and energy efficient communication has been largely unexplored. The IEEE 802.15.4 UWB physical layer allows for several settings to be selected that influence the energy consumption, range and reliability. Combined with the available link state diagnostics reported by UWB devices, there is an opportunity to dynamically select PHY settings based on the environment. To address this, we propose a deep Q-learning approach for enabling reliable UWB communication, maximizing packet reception rate (PRR) and minimizing energy consumption. Deep Q-learning is a good fit for this problem, as it is an inherently adaptive algorithm that responds to the environment. Validation in a realistic office environment showed that the algorithm outperforms  traditional Q-learning, linear search and using a fixed PHY layer. We found that deep Q-learning achieves a higher average PRR and reduces the ranging error while using only 14\% of the energy compared to a fixed PHY setting in a dynamic office environment. 
\end{abstract}


\IEEEpeerreviewmaketitle

\section{Introduction}
\IEEEPARstart{U}{ltra-wideband} (UWB) technology has recently attracted the attention of not only the research community, but also the public. This is due to the addition of the technology to several smartphones and other consumer products \cite{coppens2022overview}. UWB uses a much wider bandwidth i.e., more than 500 MHz, compared to traditional narrowband techniques such as Narrowband Internet of things (NB-IoT), Sigfox, LoRa, etc. This has benefits such as higher robustness to multipath effects, high temporal resolution and high channel capacity \cite{Alarifi2016}. The high temporal resolution allows UWB to be used for precision ranging and enables applications like hands-free access-control and indoor navigation. While UWB is more robust to multipath effects (compared to narrowband techniques), the performance is still highly dependent on the environment. Having Line-of-sight (LOS) or Non-line-of-sight (NLOS) conditions and destructive interference influence the calculated ranges and reliability dramatically \cite{NLOSeffect}. This has led to a large quantity of work to develop complex algorithms in order to maximize the ranging accuracy of UWB in these situations. Several approaches for this have been explored, such as auto-encoders \cite{fontaine2020edge}, probabilistic learning \cite{problearning} and path detection algorithms \cite{pathdetection}.

The IEEE 802.15.4 UWB physical layer (PHY) standard \cite{ieeestd} contains several settings that can be selected: for example, channel, data rate (DR) and preamble symbol repetitions (PSR). These settings have shown to be highly influential on the radio sensitivity and energy consumption \cite{uwbruntimeadap} and could enable more reliable and energy efficient UWB communication. However, it is challenging to find a reliable setting when a poor link is detected. For example, multipath reflections can cause destructive interference and to avoid it, a change of channel (frequency) is necessary even if it would be in theory a less reliable channel. Currently, hard-coded UWB PHY settings are used in most scientific papers and practical UWB systems, and these are not dynamically adapted with respect to a change in the environment \cite{uwbruntimeadap}. This makes them ineffective at sustaining a high reception rate \cite{polypoint}. A second factor is that Qorvo DW1000-based devices report numerous link quality measures for each received packet, such as the received (RX) power, first path (FP) power, channel impulse response (CIR) and so on \cite{DW1000man}. This information could be used to characterize the environment and aid in selecting the optimal PHY settings. The combination of the influence of the PHY layer settings and the available link state information gives the opportunity to dynamically select PHY settings based on the environment. By doing this separately on different UWB links, we can enable a decentralized solution for adaptive UWB communication in complex, dynamic environments without the need for a central gateway to handle all communication. This dynamic link adaption is still mostly unexplored for UWB communication in scientific literature, certainly compared to accuracy improvement algorithms.

Since the aim is to solve the problem in a decentralized way, i.e., directly on anchors, reinforcement learning (RL) is well suited as it can reach an optimal solution for a control problem by interaction with the environment. A RL agent interacts with the environment in discrete time steps. At each time, it takes an action and then receives feedback for that action in the form of a reward. The agent uses a policy to maximize the expected cumulative rewards by rewarding desired behavior and punishing undesired ones. In RL, the environment can be formulated as a Markov Decision Process (MDP), the value of the current state should suffice to determine transition probabilities and rewards following an action selection. This enables it to perceive and interpret its environment without requiring a model of the environment. Q-learning \cite{watkins1992q} is a commonly used RL algorithm that has already been used for several cognitive communication applications \cite{qcognitive1,qcognitive2, shahid2015docitive, girmay2021coexistence, maglogiannis2018q}. However, Q-learning is only suitable for problems with a low-dimensional state space because the learned policy is stored in a table. This is unfeasible for large action and state spaces, as it a) lacks generalization \cite{6gdrl} and b) increases the size of the Q-table, which would result in slower convergence. To mitigate this, the table can be replaced by a neural network that takes the state as input and approximates the Q-table values. This method is called deep Q-learning, and it has also shown its potential in cognitive communication applications \cite{RLV2V,dqnci,6gdrl}.
\\
\\
The main contributions of the paper are the following:
\begin{itemize}
    \item Design of a decentralized approach for solving adaptive UWB settings using Deep Q- and Q-learning that is executed at the anchor level without the need of information from the tag.
    \item Identification of the most relevant UWB link features for adaptive UWB PHY problems.
    \item Comparison of deep Q-learning, Q-learning and linear search in terms of PRR and energy consumption in static and dynamic environments. Algorithmic complexity and time complexity of the proposed approaches are also presented.  
     \item Collection of UWB dataset in a realistic environment using 72 different UWB PHY settings, which is made publicly available\footnote{The dataset will be made available together with the final version of the paper, and is currently available on request.  https://github.com/dietercoppens/UWB-DRL-PHY-Runtime-Adaptation-dataset}.
\end{itemize}

The remainder of this paper is organized as follows. Section \ref{sec:relatedwork} discusses the related work for both UWB PHY adaptation and deep Q-learning. In section \ref{sec:problem}, the UWB communication problem and the system model are described. Section \ref{sec:algoritmdesign} describes the proposed deep Q-learning and Q-learning algorithms. Next, sections \ref{sec:officelab} and \ref{sec:dataset} describe the environment in which the dataset is gathered and how the measurements are performed. Section \ref{sec:results} discusses the performance of the developed algorithms in a static and dynamic environments. This is followed by the conclusion in Section \ref{sec:conclusion}
\section{Related Work}
\label{sec:relatedwork}
The related work is divided into a) adaptive UWB communication and b) Q-learning and Deep Q-learning for wireless communication. 
\subsection{Adaptive UWB communication}
Although numerous papers aim to optimize the ranging accuracy by using advanced algorithms \cite{problearning,pathdetection,fontaine2020edge}, e.g. by performing error correction or detecting LOS or NLOS, these solutions typically do not consider link reliability or energy consumption aspects. This section provides an overview of recent papers that investigate the influence UWB PHY settings on these metrics. 

The authors of \cite{uwbruntimeadap} present a study of the influence of different PHY settings on the reliability and energy efficiency of UWB communication. For example, increasing the pulse repetition frequency (PRF) provides a slight improvement in reliability, but it reduces the energy efficiency and increasing the PSR provides a greater increase in reliability, but it also has a worse effect on the energy efficiency.  This study shows that by adjusting these settings, the radio sensitivity can be increased, but this generally comes at the cost of energy efficiency. Next, they estimate the link quality based on the CIR and try to extract characteristics of the surrounding environment from this. Finally, a scheme is designed to adapt the PHY setting based on the estimated link quality. The only used link state metric is CIR, and the whole logic is based on experimental data in only LOS conditions, which is not realistic for practical applications where NLOS conditions cannot be ignored. For characterization of LOS/NLOS situations to adapt the PHY setting, the authors refer to other papers using machine learning techniques. 
This paper aims to solve these limitations by using several metrics to characterize the environment, using realistic experimental data in both LOS and NLOS situations and removing the need for separate algorithms to detect LOS/NLOS. In \cite{uwbphypower}, the energy consumption of UWB communication is analyzed based on the PHY layer setting. The authors show that more energy is consumed for lower data rates due to the higher number of pulses that need to be transmitted, which causes the system to be turned on longer. \cite{uwbranginganalysis} focuses on the influence of the UWB PHY settings on the ranging accuracy when using asynchronous two-way ranging. The results were obtained using an extensive measurement campaign where more than 200 UWB PHY settings were tested. The used radio channel, DR and PRF were found to strongly influence the ranging accuracy, but no algorithm was provided to dynamically modify the setting based on the environment. The authors of \cite{uwberrors} present a method to give a reliability indication to an UWB distance measurement. For this, diagnostic information available on Qorvo's DW1000 UWB radio chip is used. The diagnostic information, such as CIR and leading edge detection (LDE)—threshold, are processed together to determine the estimated quality of the ranging measurement. In \cite{uwbphyadapt2}, a framework to make UWB more robust is proposed. The framework uses the distance between the first path and the ambient noise to determine low quality ranging. If low quality ranging is detected, a linear search algorithm that tries out all UWB settings within certain data rate, energy consumption, error rates, and robustness requirements. The best performing setting is then selected. This approach has some drawbacks, the UWB settings are only changed when a certain quality threshold is passed, which means that the performance is not continuously evaluated and optimized and the linear search algorithm means that every possible UWB PHY settings needs to be tried out first before the best performing one is found and configured. 

The authors of these papers discuss the influence of the PHY settings on the reliability and energy consumption and how to give a quality indication to a range. However, they do not elaborate on how to use the link state information and link diagnostics to dynamically improve the UWB reliability and energy consumption of UWB communication. We will address these limitations by developing a RL algorithm that dynamically selects the best performing setting without a) the need to try out every possible setting and b) requiring additional knowledge of the link state parameters. The best performing setting is continuously chosen by the algorithm.

\subsection{Q-learning and Deep Q-learning for wireless communication}
Q-learning and Deep Q-learning have been used for several resource management wireless communication problems. The authors of \cite{RLV2V} proposed Deep Q-learning for a resource allocation mechanism for V2V communications. In \cite{6gdrl} it is used for downlink power allocation in multicell systems and the authors of \cite{dqnci} proposed Deep Q-learning for circumstance-independent resource allocation with efficient scheduling and power allocation. These papers show that RL, and in particular Q- and deep Q-learning, can be successful in wireless communication resource management. However, it has not yet been used for UWB communication, as it requires a different system and algorithm design and the proposed algorithms in wireless communication resource management problems cannot be directly applied for solving the adaptation of UWB PHY settings problem.
There is a clear research gap in (1) developing a decentralized algorithm that adapts the UWB PHY setting in runtime using deep Q-learning while taking into account both reliability (PRR) and energy consumption and (2) evaluating the algorithm on a realistic dataset containing both LOS and NLOS situations.
\section{Problem and system description}
\label{Problem and system description}
\begin{table*}
\centering
\caption{List of DW1000 provided link state information parameters}
\label{tab:DW1000linkstateparam}
\begin{center}
\begin{tabular}{|l|l|l| }
\hline
\textbf{Features (Abbreviation)} & \textbf{Description}  \\
\hline
$CIR$    & The array of accumulated complex Channel Impulse Response (CIR) \\
\hline
$F_1$    & The amplitude of the first harmonic of the first path (FP) signal \\
\hline
$F_2$    & The amplitude of the second harmonic of the FP signal \\
\hline
$F_3$    & The amplitude of the third harmonic of the FP signal \\
\hline
$CIR\_{power}$    & The CIR power\\
\hline
$N_{c}$      & The standard deviation of the noise reported in the CIR accumulator \\
\hline
$RX_{pacc}$          & The preamble accumulation count at the receiver \\
\hline
$fpindex$         & The index of the detected FP \\
\hline
$FP_p$       & The estimated FP power level in $dBm$ and can be calculated as\\
                  & \textrm{FP power level}$=10\log_{10}\left(\frac{F_1^2+F_2^2+F_3^2}{N^2}\right)-A$ \\
                  & where $A$ is $113.77$ for a PRF of $16 MHz$ and 121.74 for a PRF of $64 MHz$, \\
                  & $N$ represents preamble accumulation count\\
\hline
$RX_p$       & The estimated RX power level in $dBm$ and can be calculated as\\
                  &\textrm{RX power level} = $10 \log_{10}\left(\frac{CIR_{power}\times 2^{17}}{N^2}\right)-A$\\
\hline
$NLOS$           & The power difference between the RX power and FP power and can be calculated as\\
                  & $\textrm{NLOS} = \textrm{RX power level} - \textrm{FP power level}$\\
                  & Can be used as an indicator for NLOS. If greater than 10 dB, likely to be in NLOS \\
\hline
$PR$           & The power ratio between the RX power and FP power and can be calculated as\\
                  & $\textrm{PR} = \textrm{RX power level} / \textrm{FP power level}$\\
\hline
$\hat{d}$      & The estimated range and can be calculated as $\hat{d}=c\times\tau$ \\
                              & where $c$ represents the speed of light in $m/s$ and $\tau$ the signal propagation time from tag to anchor.\\
\hline
$LDE$    & Leading edge detection (LDE) threshold used to find the first path, based on an estimate of the noise. \\ 
\hline
$PP$    & Amplitude of the peak path (PP) in the CIR. \\ 
\hline
$PP\_index$    & Index of the PP in the CIR. \\ 
\hline
$Q_1$    & The ratio between $F_2$ and $N_c$   \\
         & $Q_1 = F_2 / N_c $\\
         & Comparing the noise with $F_2$ can give additional indication of the quality of the FP measurement.\\
\hline
$Q_2$    & The ratio between $F_2$ and $LDE\_threshold$   \\
         & $Q_1 = F_2 / LDE\_threshold $\\
         & Comparing the $LDE\_threshold$ with $F_2$ gives additional indication of the quality of the $LDE\_threshold$ \\
\hline
\end{tabular}
\end{center}
\end{table*}
\label{sec:problem}

\begin{figure}[ht]
    \centering
    \includegraphics[width=0.45\textwidth]{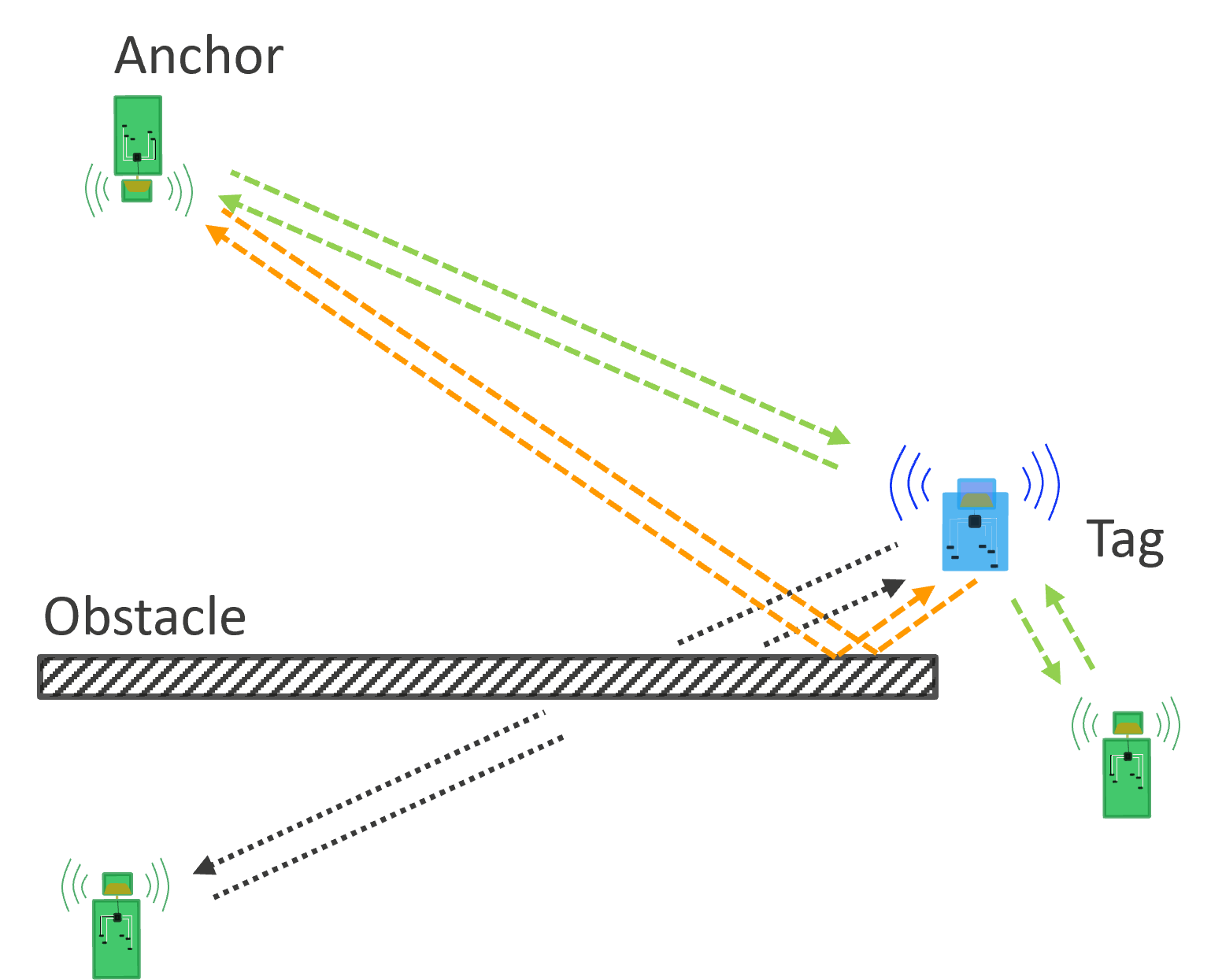}
    \caption{An illustrative picture of a complex indoor environment causing multipath and attenuated communication}
    \label{fig:uwbenvironment}
\end{figure}

Figure \ref{fig:uwbenvironment} illustrates environmental challenges that cause problems for UWB communication in indoor environments. Obstacles in the environment can cause attenuated NLOS paths between the devices, and reflections can cause multipath and interference. These effects have a major impact on both the reliability and ranging accuracy.

\subsection{UWB link state information}
The Qorvo DW1000 UWB radio chip reports several diagnostic link state parameters that can be used to characterize the environment and determine the state of the link. Table \ref{tab:DW1000linkstateparam} gives an overview of the link state parameters that are available. All diagnostics can be found in the user manual \cite{dw1000manual} as well. Using these parameters, estimates can be made of link quality \cite{uwbruntimeadap,uwberrors} and the presence of LOS/NLOS \cite{nlosdetect,DW1000man}.

   \begin{figure}[h]
    \centering
    \includegraphics[width=0.48\textwidth]{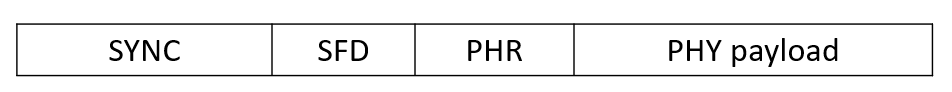}
    \caption{Structure of a UWB PHY frame in the IEEE 802.15.4 standard \cite{ieeestd}.}
    \label{fig:structure4}
\end{figure}

\subsection{UWB PHY settings}
In UWB communication, several PHY settings can be configured that have an influence on the sensitivity and energy consumption. Important to mention is that in this section, we focus on the settings that are supported by the Qorvo DW1000 chip that is used to gather the dataset.
\begin{itemize}
    \item \textbf{Channel}: The IEEE 802.15.4 UWB PHY defines 16 different channels. A channel is a combination of center frequency and bandwidth. Changing the center frequency and bandwidth influences the perceived noise and can be an important parameter to increase the robustness.
    \item \textbf{PSR}: The UWB PHY packet (shown in Figure \ref{fig:structure4}) starts with a synchronization header containing a preamble (used for frame synchronization and detection) and a start-of-frame delimiter (SFD) that indicates the end of the preamble and the time of arrival used for ranging. The length of the preamble can be changed by changing the amount of repetitions of the preamble symbol that are sent. A higher PSR increases the robustness because there is a higher chance of reception, but it increases the radio's energy expenditure due to the longer length.
    \item \textbf{PRF}: In UWB communication, a bit is transmitted using a train of pulses. The speed at which these pulses are sent is a configurable setting. By increasing the PRF, the amount of pulses sent per unit of time increases. This means that there are more threshold decision events during, and thus a higher accuracy and robustness. This comes at the cost of higher energy consumption.
    \item \textbf{Data rate (DR)}: Lowering the DR increases the reliability of the communication, but causes a longer transmission time and thus higher energy consumption.
    \item \textbf{Transmit power gain ($P_{tx}$)}: The transmit power gain can be adjusted, a higher power gain means reliable communication can be achieved for longer distance between the devices, but this also comes at the cost of higher energy consumption.
\end{itemize}

Summarizing the information from these settings, we can see that there is a trade-off between the reliability and energy consumption \cite{uwbruntimeadap}. However, it is challenging to find a reliable PHY setting when a poor link is detected. 
For example, multipath reflections can cause destructive interference and to avoid this, a change of channel (carrier frequency) is necessary even if it would be in theory a less reliable channel (higher carrier frequency). 

\subsection{System description}
The list of notations used in the paper is given in Table \ref{tab:notations}. We consider a link between two UWB devices, a tag and an anchor (as shown in Figure \ref{fig:uwbenvironment}) using the IEEE 802.15.4 UWB PHY standard. The goal is to evaluate the link state and adjust the PHY settings (in real time) on the anchor to ensure that the UWB communication is always as good as possible, in terms of reliability (PRR) and energy consumption, even in dynamic and changing environments. This PHY setting is then communicated to the tag UWB device and finally configured on both. The PHY setting can be combined into a single settings variable $A$ as shown in (\ref{eq:setting}).

\begin{equation}
       A =\left\{C, PSR, PRF, DR, P_{tx} \right\},
       \label{eq:setting}
\end{equation}

\noindent where $C \in \{3,5,7\}$, $PSR \in \{128, 1024, 4096\}$, $PRF \in \{16, 64\}$, $DR \in \{110, 6800\}$ and $P_{tx} \in \{0, 10.5\}$. These values are determined by our collected dataset. 

\begin{table}[]
\centering
\caption{List of notations and abbreviations used in the paper.}
\label{tab:notations}
\begin{tabular}{|ll|}
\hline
\multicolumn{1}{|l|}{\textbf{Notation/Abbreviation}} & \textbf{Description} \\ \hline
$e$ & Environment \\
C & Channel \\
DR & Data rate \\
$A$ & UWB PHY setting variable\\
TX & Transmit \\
RX & Receive \\
$P$ & Power consumption \\
$I$ & current consumption \\
$S_{p}$ & The number of symbols in the complete\\ 
        & preamble\\
$S_{s}$ & The number of symbols in the SFD\\
$S_{d}$ & The number of symbols in the data\\ 
        &part of the frame\\
$b_{p}$  & The number of bits in the PHR\\
$B_{p}$ & The number of bytes in the payload\\
$R_{FEC}$ & The forward error correction rate \\

$T_{sym}$ & The duration of a symbol\\
$T_{p}$ & The total duration of the preamble \\
$T_{d}$ & The total duration of the data part \\
$E_{p}$ & Energy consumption of the preamble \\
$E_{d}$ & Energy consumption of the data part \\
$E_{rx}$ & Energy consumption during transmission \\
$E_{tx}$ & Energy consumption during reception \\
$P_{p}$ & Power consumption during the preamble \\
$P_{d}$ & Power consumption during the data part \\
$E$ & The total energy consumption for a range\\
$G$ & The complete system with respect to $A$-th \\
    &setting and $e$-th environment \\
$s_t$ & State at time t \\
$f$ & List of all available features \\
$Xcor$ & The cross correlation between two vectors. \\
$\alpha$ & Learning rate in Bellman equation \\
$\gamma$ & Discount factor in Bellman equation \\
$F$ & $F$-value of a feature \\
$R_t$ & The reward at time $t$ \\
$\epsilon$ & The epsilon-greedy parameter \\
$\theta$ & Weights of main Deep Q-network \\
$\hat{\theta}$ & Weights of target Deep Q-network \\
$\zeta$ & Importance sampling prioritization \\
        & parameter \\
$\beta$ & Importance sampling bias compensation \\
        &parameter \\
$p_i$ & Priority or TD-error of experience $i$\\
$Prob(i)$ & Sampling probability of experience $i$ \\
$w_i$ & Importance sampling weight of experience $i$ \\
\hline
\end{tabular}
\end{table}

\begin{table*}[]
\centering
\caption{Overview of I($A$) for the different modes in the system \cite{DW1000man}}
\label{tab:currentoverview}
\begin{tabular}{|c|c|c|c|c|c|c|}
\hline
\textbf{Channel} & \textbf{\begin{tabular}[c]{@{}c@{}}PRF\\ (MHz)\end{tabular}} & \textbf{\begin{tabular}[c]{@{}c@{}}Data rate\\ (Mbps)\end{tabular}} & \textbf{\begin{tabular}[c]{@{}c@{}}Preamble\\ TX \\ current (mA)\end{tabular}} & \textbf{\begin{tabular}[c]{@{}c@{}}Preamble \\ RX \\ current (mA)\end{tabular}} & \textbf{\begin{tabular}[c]{@{}c@{}}Data\\ TX  \\ current (mA)\end{tabular}} & \textbf{\begin{tabular}[c]{@{}c@{}}Data\\ RX \\ current (mA)\end{tabular}} \\ \hline
3 & 16 & 0.11 & 68 & 113 & 35 & 59 \\ \hline
3 & 16 & 6.8 & 68 & 113 & 50 & 118 \\ \hline
3 & 64 & 0.11 & 83 & 113 & 40 & 72 \\ \hline
3 & 64 & 6.8 & 83 & 113 & 52 & 118 \\ \hline
5 & 16 & 0.11 & 74 & 118 & 42 & 62 \\ \hline
5 & 16 & 6.8 & 74 & 118 & 57 & 123 \\ \hline
5 & 64 & 0.11 & 89 & 118 & 46 & 75 \\ \hline
5 & 64 & 6.8 & 89 & 118 & 59 & 123 \\ \hline
7 & 16 & 0.11 & 74 & 118 & 42 & 62 \\ \hline
7 & 16 & 6.8 & 74 & 118 & 57 & 123 \\ \hline
7 & 64 & 0.11 & 95 & 124 & 52 & 81 \\ \hline
7 & 64 & 6.8 & 95 & 124 & 65 & 129 \\ \hline
\end{tabular}
\end{table*}
The environment is described by parameter $e$, this indicates the influence of the environment on the system. The parameter is adjusted when changes happen in the environment in which the UWB ranging is taking place. The ranging method used in the system is called Asymmetric Double Sided TWR (ADS-TWR) \cite{ADSTWR}, which means that for one range estimation, there are three packets transmitted (TX) and received (RX). Combining this with the configured setting $A$, the energy consumption can be determined. Table \ref{tab:currentoverview} shows the current consumption ($I$) for the different settings during TX and RX for both the preamble and data parts in mA. Using this information and the constant supply voltage of 3.3 V, the power $P$ during the preamble and data parts of the frame for both TX and RX can be calculated using (\ref{eq:powereqs}). Note that we use $P(A)$ which means the power with respect to the PHY setting $A$. This notation is used for other quantities as well. 
\begin{equation}
    P(A) = 3.3 \: \cdot I(A).
    \label{eq:powereqs}
\end{equation}
The number of symbols in the preamble $S_{p}(A)$ and data parts $ S_{d}(A)$ can be calculated by (\ref{eq:preamblesym}) and (\ref{eq:datasym}), using the 12 data bytes ($B_p$), the number of bits in the PHR ($b_p$) the forward error correction rate ($R_{FEC} = 0.87$) and the SFD symbols ($S_s$) (64 for 110 kbps and 8 for other data rates).
\begin{table}[]
\centering
\caption{DW1000 symbol durations in the separate parts of the frame \cite{DW1000man}}
\label{tab:symboldurations}
\begin{tabular}{|c|c|c|c|}
\hline
\textbf{\begin{tabular}[c]{@{}c@{}}PRF\\ (MHz)\end{tabular}} & \textbf{\begin{tabular}[c]{@{}c@{}}Data Rate\\  (Mbps)\end{tabular}} & \textbf{\begin{tabular}[c]{@{}c@{}} $\boldsymbol{T_{sym}}$ \\ SHR (ns)\end{tabular}} & \textbf{\begin{tabular}[c]{@{}c@{}} $\boldsymbol{T_{sym}}$ \\  Data \& PHR\\  (ns)\end{tabular}} \\ \hline
16 & 0.11 & 993.59 & 8205.13 \\ \hline
16 & 6.8 & 993.59 & 1025.64 \\ \hline
64 & 0.11 & 1017.63 & 8205.13 \\ \hline
64 & 6.8 & 1017.63 & 128.12 \\ \hline
\end{tabular}
\end{table}
\begin{equation}
    S_{p}(A) = PSR + S_{s}(A).
    \label{eq:preamblesym}
\end{equation}
\begin{equation}
    S_{d}(A) = (b_{p} + \frac{B_{p}\cdot 8}{R_{FEC}}).
    \label{eq:datasym}
\end{equation}
The duration of the preamble $T_{p}(A)$ and data $T_{d}(A)$ parts (in seconds) can then be calculated using (\ref{eq:durations1}) and (\ref{eq:durations}) where the symbol duration ($T_{sym}$) is computed from Table \ref{tab:symboldurations}.
\begin{equation}
    T_{p}(A) = S_{p}(A) \cdot T_{sym}(A).
    \label{eq:durations1}
\end{equation}
\begin{equation}
    T_{d}(A) = S_{d}(A) \cdot T_{sym}(A).
    \label{eq:durations}
\end{equation}
Using the power and duration of the separate parts, we can calculate the energy consumption of the preamble $E_{p}(A)$ and data $E_{d}(A)$ parts as shown in (\ref{eq:energyconsumption}) and (\ref{eq:energyconsumption1}). 
\begin{equation}
    E_{p}(A) = P_{p}(A) \cdot T_{p}(A). 
    \label{eq:energyconsumption}
\end{equation}
\begin{equation}
    E_{d}(A) = P_{d}(A) \cdot T_{d}(A).
    \label{eq:energyconsumption1}
\end{equation}
The energy consumption is the combination of the data and preamble parts as shown in (\ref{eq:energyparts}). 
\begin{equation}
    E(A) = E_{p}(A) + E_{d}(A). \:
    \label{eq:energyparts}
\end{equation}

Finally, the total energy consumption (in Joules) for a range using a certain setting is found using (\ref{eq:powertotal}). For the energy consumption during TX, an extra factor $10^{\frac{P_{tx}}{10}}$ is added to account for the $P_{tx}$ gain in the setting. 
\begin{equation}
    E(A) = 3 \cdot (E_{rx}(A) + E_{tx}(A)\cdot10^{\frac{P_{tx}}{10}}).
    \label{eq:powertotal}
\end{equation}

The second factor to optimize in the system is the PRR, this factor depends on both the environment $e$ and used setting A. The complete system $G(A,e)$ with respect to the $A$-th setting and $e$-th environment can be described by (\ref{eq:system}). It is a combination of the PRR and the scaled and normalized energy consumption ($E$).
\begin{equation}
    PRR = \frac{received \: packets}{total \: transmitted \: packets}.
    \label{eq:prr}
\end{equation}

\begin{equation}
    G(A,e) = PRR(A,e) + \left(1-\frac{E(A)-min(E(A)}{max(E(A))-min(E(A))}\right).
    \label{eq:system}
\end{equation}

\subsection{Problem description}
The goal is to enable reliable UWB communication while consuming as little energy as possible with the available settings. This is depicted below.
\begin{equation*}
\begin{aligned}
 \MoveEqLeft{\text{Maximize:}}& \\
& G(A,e) = PRR(A,e) + \left(1-\frac{E(A)-min(E(A)}{max(E(A))-min(E(A)}\right) \\
 \MoveEqLeft{\text{Constrained by:}} &\\
&  A
\end{aligned}
\end{equation*}

\section{Algorithm design}
\label{sec:algoritmdesign}
In this section, the proposed deep Q-learning algorithm for runtime adaptation of UWB PHY settings while maximizing $G(A,e)$ as introduced in Section \ref{Problem and system description}. First, the key parts of RL are introduced. Then, the feature selection step is shown. Next, we will discuss why the discrete nature of Q-learning limits to what extent link estimation can be optimized. Finally, a more advanced deep Q-learning algorithm is proposed. 
\subsection{Reinforcement learning}

\begin{figure}[ht]
    \centering
    \includegraphics[width=0.49\textwidth]{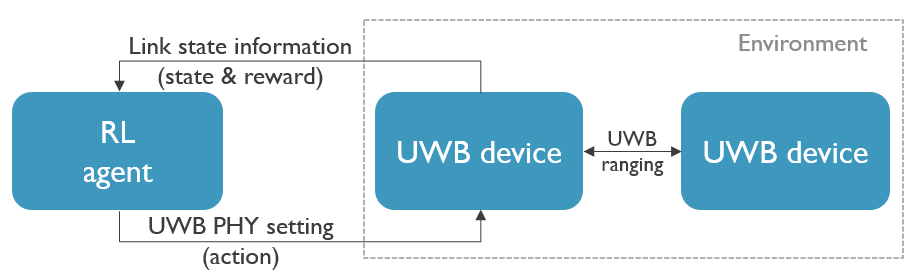}
    \caption{High-level overview of RL system for UWB PHY run-time adaptation}
    \label{fig:systemoverview}
\end{figure}

As shown in Figure \ref{fig:systemoverview}, the RL framework is composed of an agent and an environment interacting with each other. An agent considers the link between an anchor and a tag UWB device and is executed at the anchor. Anything in the area around the two UWB devices that could possibly affect communications is regarded as the environment.
As in Figure \ref{fig:systemoverview}, at each time $t$, the UWB link observes a state $s_t$, and takes an action $a_t$. The action selection is determined by a state-action function, $Q(s_t,a_t)$. The environment transitions to the next state $s_{t+1}$ and receives a reward $R_t$ based on the action taken. 
Specifically for this problem, the objective function $G$ can be seen as a first approximation of the reward. The UWB PHY setting, expressed as:
\begin{equation}
   A =\left\{C, PSR, PRF, DR, P_{tx}\right\} \nonumber,
\end{equation}

\noindent is the action space from which action $a_t$ is selected. The state is a combination of the link state parameters from Table \ref{tab:DW1000linkstateparam}, expressed as:
\begin{equation}
    s_t = \left\{CIR, F_1, F_2 ..., Q_2 \right\} \nonumber.
\end{equation}

\subsection{State feature selection}
\label{sec:features}
The DW1000 UWB chip reports a significant amount of diagnostics for assessing the link quality. Using all of them for the state $s_t$ would considerably increase the complexity of the proposed Q-learning and deep Q-learning algorithms. Therefore, a feature selection step is necessary. Reinforcement learning consists of two phases, namely an initial exploration phase, where many actions are tried out to find out in general which actions perform well and the optimization phase, where the system tries to find the optimal action to select. To select features that are representative for both phases, there are two factors based upon which features can be prioritized: (1) for exploration, their impact on selecting which action to take and (2) to find the optimal action, their influence on the objective function $G$. To reflect these two factors, feature selection is performed for both classifying which setting to select and predicting the value of $G$. To select the best features to predict $G$, we use the F-value. This calculation is performed in two steps:
\begin{figure*}
    \centering
    \includegraphics[width=0.85\textwidth]{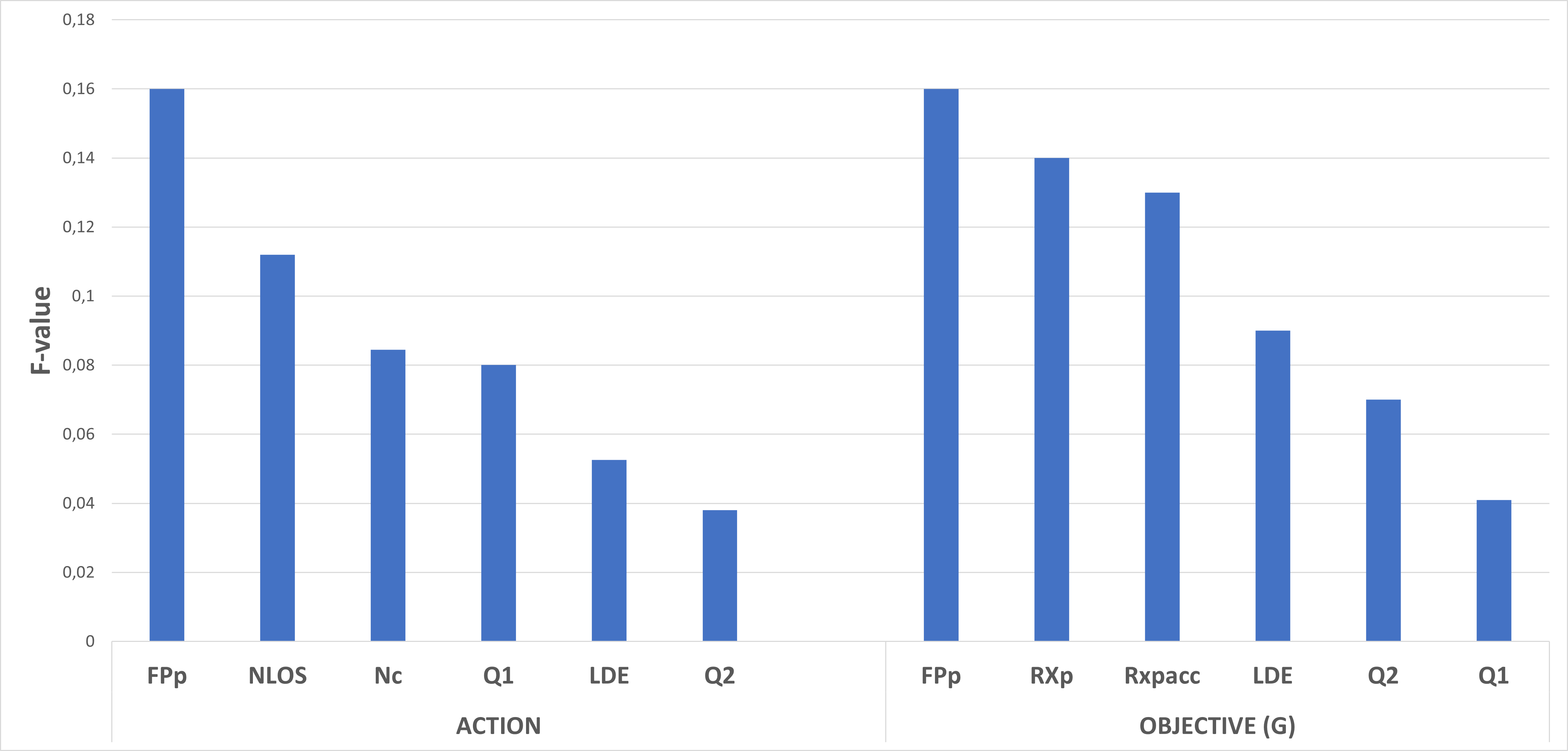}
    \caption{Overview of the most important features, based on F-value, for both classification of the setting (action) and prediction of the G used in the run-time adaptation of the UWB PHY layer}
    \label{fig:feature-overview}
\end{figure*}
\begin{enumerate}
    \item The cross-correlation $Xcor$ between each feature $f(i)$ in list of features f and target $G$ is calculated using (\ref{eq:croscor}).
    \item The cross correlation is converted to the F-value using equation \ref{eq:feq}.
\end{enumerate}

\begin{equation}
      Xcor(i) = \frac{E \left[(f[i] - mean(f[i])\cdot(G - mean(G))\right]}{std(f[i]) \cdot std(G)} 
      \label{eq:croscor}
\end{equation}
\begin{equation}
    F(i) = \frac{Xcor(i)^{2}}{1 - Xcor(i)^{2}} \cdot \nu 
    \label{eq:feq},
\end{equation}

\noindent where $\nu$ is the degrees of freedom. ($\nu = n-1$ with $n$ is the number of features $f$ in the dataset).
The features with the highest F-value have the highest correlation to the value of $G$ and will thus be the most important features. To select the best features to classify which setting is the best, the Analysis of variance (ANOVA) F-value can be used. This F-value is similar to the previously explained F-value, only now the result is a categorical value (the setting) instead of a numerical value. The F-values for the six most important features for both F-value calculations are shown in Figure \ref{fig:feature-overview}. From this, we can select the most important features to be used in our system: $RX_p$ (The estimated RX power level), $FP_p$ (The estimated FP power level), $NLOS$ (The difference between $RX_p$ and $FP_p$), $N_c$ (The standard deviation of the noise), $Q_1$ (The ratio between the amplitude of the second harmonic ($F_2$) and $N_c$), $RX_{pacc}$ (The preamble accumulation count), $LDE$ (leading edge detection threshold, $Q_2$ (The ratio between the amplitude of the second harmonic ($F_2$) and $LDE$, PRR (The ratio of received packets to total transmitted packets).

\noindent These will be used as the state of the RL system, expressed as:
\begin{equation}
    s_t = \left\{RX_p, FP_p, NLOS, N_{c}, Q_1, RX_{pacc}. LDE, Q_2,PRR \right\} \nonumber,
\end{equation}

\subsection{Q-learning}
\label{sec:qlearning}
\begin{figure}[ht]
    \centering
    \includegraphics[width=0.48\textwidth]{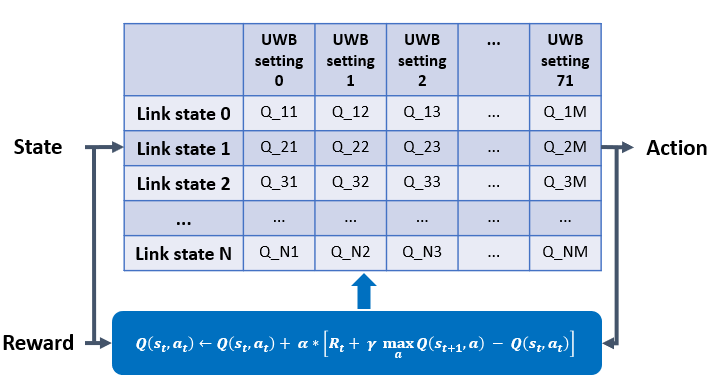}
    \caption{Schematic showing the Q-table and Bellman update equation for the Q-learning run-time adaptation of UWB PHY settings}
    \label{fig:Q-detailed}
\end{figure}

Q-learning is a model-free RL algorithm that learns the value of an action in a certain state. The most important part of the algorithm is the Q-table. Each row in the table represents a state of the system, and each column represents an action. Each value in the table represents the 'quality' of a particular state-action pair. A more detailed schematic of the Q-learning algorithm is given in Figure \ref{fig:Q-detailed}.

\subsubsection{Updating the Q-table}
The foundation of the Q-learning algorithm is the Bellman equation that is a value update function using the newly received information and weighted old value:
\begin{equation}
\begin{aligned}
    Q_{new}(s_{t},a_{t}) &\leftarrow Q_{old}(s_{t},a_{t}) \\
    &+ \: \alpha \left[R_{t} + \gamma\max_{a}Q(s_{t+1},a_{t}) - Q_{old}(s_{t},a_{t})\right].
\end{aligned}
\end{equation}
The following parameters are used in the equation:
\begin{itemize}
    \item $\alpha$: learning rate. This factor determines the weight that is given to the newly acquired information and how much old information can be overridden.
    \item $\gamma$: discount factor. This factor determines the weight that is given to newly acquired information.
\end{itemize}
Using this function, the values in the Q-table are filled in. After enough iterations, the values in the table will reflect the quality of the state-action pairs and the expected value of the total received rewards will be maximized. The reward clearly drives the behavior of the algorithm and the reward function, a mapping of a state-action pair to a numerical value that indicates the desirability, is thus crucial. Applied to this problem, the reward function needs to describe the value of selecting a certain setting based on the link state information. The objective function $G$ discussed before is a potential reward function. However, this function has a problem. It could lead to the selection of a certain setting based purely on a low energy consumption. Having a low energy consumption is meaningless when no communication is possible using that low energy setting. To mitigate this problem, the following reward function is proposed:
\begin{equation}
    R_{t} = PRR + PRR\cdot \left(1-\frac{E-min(E)}{max(E)-min(E)}\right)
    \label{eq:rewardfunction}.
\end{equation}

In this function, the energy consumption factor is multiplied by the PRR. This causes the energy consumption to influence the reward proportionally to the reliability of the communication.

At each point in time, the agent selects an action to be taken using the Q-table. This action selection is a fundamental trade-off in RL. Initially, the agent does not yet know the outcome of the possible actions. Hence, high enough exploration (not selecting action with highest Q-value) is required. Once the agent has learned more information, the exploration can be reduced by exploiting (selecting the action with highest Q-value) the learned information. However, only exploiting is a dangerous approach, as the agent can get stuck in a suboptimal state.

During the training stage, the epsilon-greedy policy is used. In this policy, the best action (highest Q-value) is selected with a probability of 1-$\epsilon$. With a probability of $\epsilon$ a random action is chosen uniformly.

For evaluation, the epsilon-greedy policy is modified. Now, in exploration, the random selection is changed to a random selection among the 10 actions with the highest Q-value.

\subsubsection{State determination}
The biggest drawback in using Q-learning is that we need to determine a discrete state number from continuous state variables. The more granularity we introduce in the states, the more states are needed and the size of the Q-table increases as well. The state consists of the PRR and the selected features, each split up in 3 categories: low, middle and high. This gives a total of 19,683 possible link states or rows in the Q-table. Multiplying this with the number of actions, there is a total of 1,417,176 cells. The Q-table is huge, while we have only introduced very limited granularity on the features and performed feature selection. Without feature selection and the same limited granularity, there would be $83.7e^9$ cells in the Q-table. This demonstrates the need for the feature selection step. The second drawback is that we also need to determine the low, middle and high separation for each feature manually. This means that some 'expert knowledge' of UWB diagnostics is necessary that has a large influence on the performance of the algorithm.

The pseudocode for the complete Q-learning algorithm is given in Algorithm \ref{alg:q}. The Q-learning algorithm was trained for 500,000 steps with $\alpha = 0.8$ and $\gamma=0.5$. $\epsilon$ (from the epsilon-greedy policy) changed during the training, following an exponential decay (\ref{eq:expdecay}).With $\lambda$ decay constant and 'step' the amount of training steps that have already been performed. 
\begin{equation}
    \epsilon = \epsilon_{min} + (\epsilon_{max} - \epsilon_{min}) \cdot e^{-\lambda\cdot step}
    \label{eq:expdecay}
\end{equation}
\noindent where $\epsilon_{min} = 0.01$, $\epsilon_{max} = 1$ and $\lambda = 3.91e^{-5}$.
\begin{algorithm}
\caption{Q-learning for UWB PHY}\label{alg:q}
\begin{algorithmic}[1]
\Require
\Statex Initialize Q-table
\Statex Initialize parameters $\alpha$, $\gamma$ and $\epsilon$ 
\Ensure
\State Select random UWB link and action (setting)
\State Determine start state $s_{t}$
\While{step $<$ training steps}
\State \textbf{Every 100 steps:} select new random UWB link
\State With probability $\epsilon$ select random action $a_{t}$
\State Otherwise  $a_{t}$ = $arg\max_{a}(Q(s, a))$
\State Perform action $a_{t}$ (configure setting)
\State Measure link quality indicators
\State Determine reward $R_{t}$ and new state $s_{t+1}$
\State $Q_{new}(s_{t},a_{t}) \leftarrow Q_{old}(s_{t},a_{t})$ 
\Statex $+ \alpha \: \left[R_{t} + \gamma \cdot \max_{a}Q(s_{t+1},a_{t}) - Q(s_{t},a_{t})\right]$
\State Update $\epsilon$
\State $s_{t}$=$s_{t+1}$
\EndWhile
\end{algorithmic}
\end{algorithm}

\subsection{Deep Q-learning}
\begin{figure}[h]
    \centering
    \includegraphics[width=0.48\textwidth]{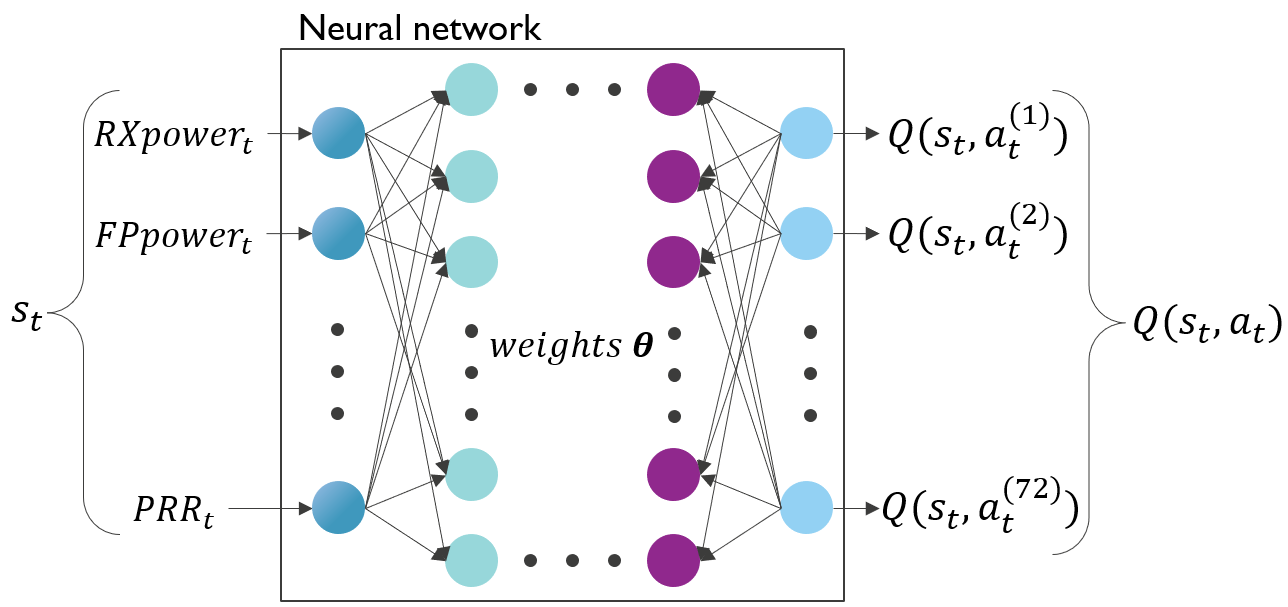}
    \caption{Illustration of input and output of a Deep Q-network}
    \label{fig:deepqnetwork}
\end{figure}

To mitigate the previously mentioned drawbacks, the Q-learning algorithm can be modified to a Deep Q-learning algorithm. In Deep Q-learning, the Q-table is replaced by a neural network that approximates the Q-value function. As illustrated in Figure \ref{fig:deepqnetwork}, the state is given as input and the Q-value of all possible actions is the output. 
Using a neural network instead of a table has several benefits:
\begin{itemize}
    \item The continuous values of the link state measurements can now be used directly as input (only scaling between 0 and 1).
    \item  No 'expert knowledge' is necessary to determine categories for the inputs.
    \item No loss of information due to discretization of measurements.
    \item No need for a table containing millions of cells.
\end{itemize}

\subsubsection{Target network}
In supervised learning, the label of an input does not change over time. This stable condition for input and output allows it to perform well. In RL, both the input and the target to obtain change constantly during the learning process. This makes training unstable because the target or 'label' of the output $Q$ depends on $Q$ itself. (\ref{eq:target}) shows the target value based on the Bellman equation with $\theta$, the weights of the $Q$ neural network.
\begin{equation}
    target = R_{t} + \gamma \cdot \max_{a}Q(s_{t+1}, :,\theta)).
    \label{eq:target}
\end{equation}
This means that the target will move as the $Q$ approximation improves. The moving target could cause the 'chasing your own tail' problem \cite{omisore2021novel}. To mitigate this issue, a target network can be introduced. The target network is used to calculate the $Q$-values and fix the target for as long as we fix the target network. The main network updates its weights $\theta$ using the training data to minimize the following loss function (\ref{eq:loss}) on a dataset $D$.
\begin{equation}
    Loss(\theta) = \sum_{s_{t},a_{t} \in D} (target-Q(s_{t},a_{t},\theta))^{2}.
    \label{eq:loss}
\end{equation}

After several updates of the main network, the weights of the main network are copied to the weights of the target network $\hat{\theta}$. In practice, we do not have two separate networks, but two instances of the neural network weights. This prevents the training process from spiraling and increases the chances of convergence \cite{mnih2015human}. The neural network architecture is given in Table \ref{tab:neuralnetwork}. The input to the neural network consists of 14 values: a) 8 state features, b) the 5 current configured PHY layer settings and c) the current PRR. The output is an approximation of the Q-values for all actions in this state. 

\begin{table}[]
\centering
\caption{Neural network architecture}
\label{tab:neuralnetwork}
\begin{tabular}{ll}
\hline
\textbf{DNN} & \textbf{} \\ \hline
\textbf{Layer} & \textbf{Output dimension} \\ \hline
Input & 14 \\
Dense(128), ReLu & 128 \\
Dense(256), ReLu & 256 \\
Dense(512), ReLu & 512 \\
Dense(256), ReLu & 256 \\
Dense(128), ReLu & 128 \\
Dense(72), linear & 72 \\
Output & 72 \\\hline
\end{tabular}
\end{table}

\subsubsection{Weighted importance sampling}
Updating and training a neural network is quite different from updating a cell in a Q-table. Updating the neural network at every time step with one sample would be very inefficient. Therefore, the network is updated on batches of data that are sampled from a replay memory containing experiences $(s_{t},a_{t},R_{t},s_{t+1})$ of the $Q$-agent. The sampling of the experiences from the memory can be done in several ways. Popular methods are random sampling, prioritized sampling (experiences with the highest reward) and weighted importance sampling \cite{weighted}. 

For this problem, we opted for weighted importance sampling. The central component of this is the criterion on which the experiences are selected. The optimal criterion would be the amount the $Q$-agent can learn from using an experience. While this measure is not available, a reasonable approximation is the temporal-difference (TD)-error, which indicates how surprising a certain transition is. However, greedy TD-error prioritization has some issues. First, to avoid expensive sweeps over the entire replay memory, TD errors are only updated for the transitions that are replayed, which means that transitions that have a low TD error on the first visit may not be replayed for a very long time (or ever). Further, greedy prioritization focuses on a small subset of the experience. This lack of diversity makes the system prone to overfitting. To overcome these issues, a stochastic sampling method that interpolates between pure greedy prioritization and uniform random sampling can be used \cite{weighted}. The priority or TD-error $p_{i}$ of an experience is calculated in (\ref{eq:priority}) and the sampling probability of each experience is given in (\ref{eq:samplingprob}).
\begin{equation}
    p_{i} = R_{t} + \gamma \cdot \max_{a}\hat{Q}(s_{t+1}, :,\hat{\theta})) - Q(s_{t},a_{t}).
    \label{eq:priority}
\end{equation}
\begin{equation}
    Prob(i) = \frac{p_{i}^{\zeta}}{\sum_{k} p_{k}^{\zeta}},
    \label{eq:samplingprob}
\end{equation}
\noindent where $\zeta$ is a hyperparameter that determines how much prioritization is used. This priority is saved with the experience. Prioritized learning in this way introduces bias, as it changes the solution that the estimates will converge to \cite{weighted}. This bias can be corrected by using importance sampling weights (w) shown in (\ref{eq:samplingweights}).
\begin{equation}
    w_{i} = \left(\frac{1}{N} \cdot \frac{1}{Prob(i)}\right)^{\beta},
    \label{eq:samplingweights}
\end{equation}
\noindent where $N$ is the amount of experiences in the memory, and $\beta$ a hyperparameter that determines how much the non-uniform probabilities $P(i)$ are compensated.

\subsubsection{Update function}
Combining this, the update function to generate the training data is given by equation (\ref{eq:totalqupdate})

\begin{equation}
    Q(s, a_{t},\theta) = Q(s, a_{t},\theta) + \alpha \cdot p_{i}\cdot w_{i}
    \label{eq:totalqupdate}
\end{equation}

The pseudocode for the Deep Q-learning algorithm is given in algorithm \ref{alg:deepq}. This algorithm was trained for 200,000 steps, with $\alpha = 0.8$, $\gamma = 0.5$. $\epsilon$ changed during training following equation \ref{eq:expdecay} with  $\epsilon_{min} = 0.01$, $\epsilon_{max} = 1$ and $\lambda = 1.96e^{-5}$.

\begin{algorithm}
\caption{Deep Q-learning for UWB PHY}\label{alg:deepq}
\begin{algorithmic}[1]
\Require
\Statex Initialize replay memory $D$ to capacity $N$
\Statex Initialize action-value function $Q$ with random weights $\theta$
\Statex Initialize target action-value function $\hat{Q}$ with $\hat{\theta}$ = $\theta$
\Ensure
\State Select random UWB link and start action (setting)
\State Determine start state $s_{t}$
\While{step $<$ training steps}
\State \textbf{Every 100 steps:} select new random UWB link
\State With probability $\epsilon$ select random action $a_{t}$
\State Otherwise  $a_{t}$ = $arg\max_{a}(Q(s, a,\theta))$
\State Perform action $a_{t}$ (configure setting)
\State Measure link quality indicators
\State Determine reward $R_{t}$, new state $s_{t+1}$ and priority $p_{t}$
\State Store  $e_{t}$ = ($s_{t}$, $a_{t}$, $R_{t}$, $s_{t+1}$) with priority $p_{t}$ in $D$
\State \textbf{Every M steps:}
\State \hspace{\algorithmicindent} Sample minibatch b of $e_{j}$ $\sim Prob(j) = \frac{p_{j}^{\zeta}}{\sum_{k} p_{k}^{\zeta}}$
\State \hspace{\algorithmicindent} For all $e_{j}$ in b: current\_Qs: $Q(s, :,\theta)$
\State \hspace{\algorithmicindent} For all $e_{j}$ in b: future\_$Q$: $\hat{Q}(s_{t+1}, :,\hat{\theta}))$
\State \hspace{\algorithmicindent} X = $s_{t}$ for all $e_{j}$ in $b$
\State \hspace{\algorithmicindent} for all $e_{j}$ in b:
\State \hspace{\algorithmicindent}\hspace{\algorithmicindent} $max\_future\_q = R_{t} + \gamma\cdot \max(future\_Qs)$
\State \hspace{\algorithmicindent}\hspace{\algorithmicindent} $p_{i} = |$max\_future\_q - current\_Qs[$a_{t}$]$|$
\State \hspace{\algorithmicindent}\hspace{\algorithmicindent} $w_{i} =\left(\frac{1}{Prob(i)} \cdot \frac{1}{N}\right)^{\beta} $
\State \hspace{\algorithmicindent}\hspace{\algorithmicindent} current\_Qs[$a_{t}$] = current\_Qs[$a_{t}$] + $\alpha \cdot p_{i}\cdot w_{i}$

\State \hspace{\algorithmicindent} $Y$ = current\_$Q$s for all $e_{j}$ in b
\State \hspace{\algorithmicindent} Update weights $\theta$: fit  $Q$ using $X$ and $Y$.
\State \textbf{Every T steps:} reset $\hat{Q}$ = $Q$, i.e., $\hat{\theta}$ = $\theta$
\EndWhile
\end{algorithmic}
\end{algorithm}

\section{Experimental evaluation}
In this section, we evaluate the proposed Q-learning and deep Q-learning algorithms to assess their performance compared to a linear search algorithms and fixed UWB PHY settings. First, the environment in which the data for the experiments is gathered is described. Then, we describe how the measurements are performed, and finally we discuss and analyze the results.
\subsection{Office lab}
\label{sec:officelab}
\begin{figure}[ht]
    \centering
    \includegraphics[width=0.48\textwidth]{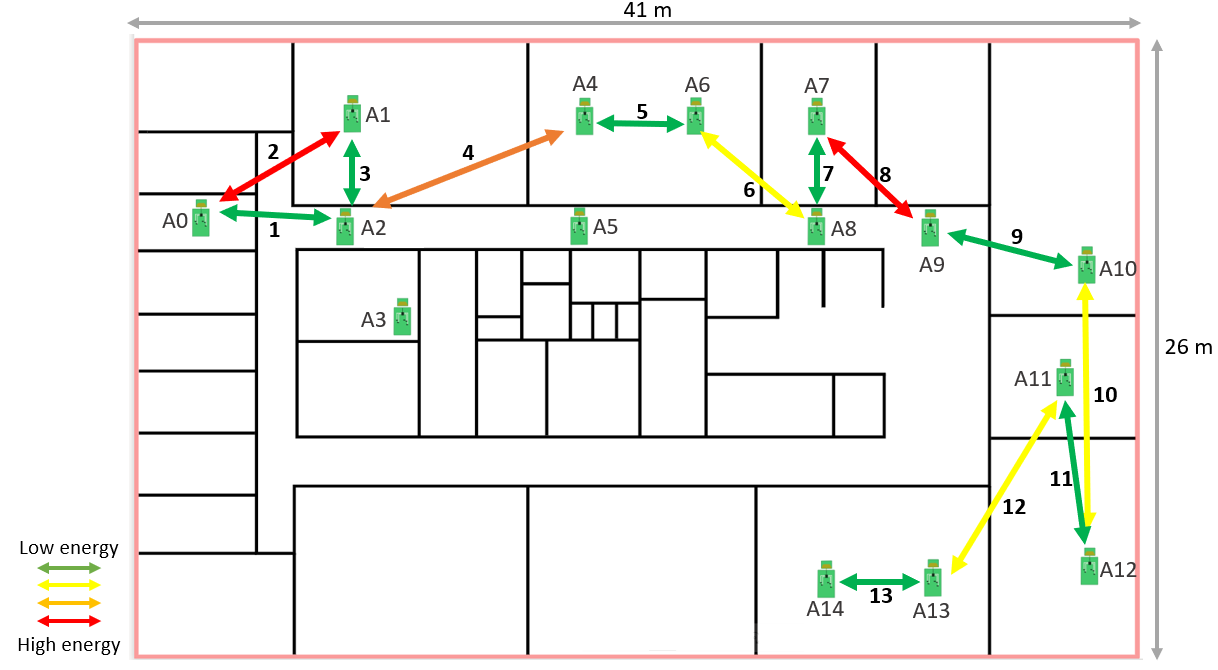}
    \caption{Data gathering setup with 15 UWB nodes (A0-A14) and the anchor-tag links selected for dynamic evaluation distributed over the 9th floor of the OfficeLab at imec - IDLab - Ghent University}
    \label{fig:9thfloor}
\end{figure}

The OfficeLab from of imec - IDLab - Ghent University \cite{officelab} offers a test environment which includes 3 floors that are equipped with 40 Intel NUC nodes, supporting several Wi-Fi and sensor technologies including UWB. In our evaluation, we use a single floor that has a total area of around 41 x 26 m$^{2}$ and 15 UWB nodes placed in corridors, meeting rooms, and offices. All nodes are placed at the same height of around 2.6 m above the floor. The walls separating the rooms consist of heterogeneous materials, ranging from plywood to reinforced concrete, resulting in a very heterogeneous environment. An overview of the placement of the nodes is shown in Figure \ref{fig:9thfloor}.

\subsection{Dataset}
\label{sec:dataset}
The dataset was gathered using Wi-PoS: a Low-Cost UWB Hardware Platform with Long Range Sub-GHz Backbone \cite{wipos}. All 15 nodes are programmed once as tag and try to range with the other 14 nodes (anchors) using the following settings shown in Table \ref{tab:UWBPHYsettingoverview}. 

\begin{table}[ht]
\centering
\caption{Parameters used in static test scenario for runtime adaptation of UWB PHY settings}
\label{tab:UWBPHYsettingoverview}
\begin{tabular}{|l|l|}
\hline
\textbf{Parameter} & \textbf{Values} \\ \hline
Channel & 3,5,7 \\ \hline
PSR & 128, 1024, 4096 \\ \hline
PRF & 16, 64 MHz \\ \hline
Data rate & 110, 6800 kbps \\ \hline
Transmit power gain & 0,10.5 dB \\ \hline
\end{tabular}
\end{table}

The combination of these settings give a total of 72 different UWB PHY settings. ADS-TWR \cite{ADSTWR} is used to enable accurate ranging. In the bidirectional communication, both nodes log their raw responses, which are then combined and converted to a more readable file format, e.g., csv. Each row of the dataset corresponds to one range between two nodes and contains the distance estimation and the available link state information that is needed in the algorithms. All devices have ranged with each other for a total of 500 range attempts per combination. This resulted in a total of more than 605 thousand ranges overall. This is considerably less than the total attempts, as on average each node could only range with 7 other nodes due to obstacles like walls.

\subsection{Results}
\label{sec:results}
In this section, we evaluate the proposed Q-learning and deep Q-learning algorithms to assess their impact on (1) the time it takes to find the best setting in a static situation (the environment and link is fixed) compared to a linear search algorithm and (2) the reliability (PRR) and energy consumption in a dynamic or changing environment compared to using fixed UWB PHY settings.
\subsubsection{Static environment} This experiment tries to evaluate the ability of the proposed model to quickly, in terms of RL iterations (or amount of settings configured), determine the best PHY layer setting when UWB communication is started between two devices. For every possible pair of UWB devices in the dataset, the models try to determine the best PHY setting in a certain amount of settings that can be configured. We compare our models with a linear search algorithm that sequentially goes through all 72 possible PHY settings and then selects the one that performed the best. The parameters of the models used during this evaluation are shown in Table \ref{tab:paramstatic}.

\begin{table}[]
\centering
\caption{Parameters used in static test scenario for runtime adaptation of UWB PHY settings}
\label{tab:paramstatic}
\begin{tabular}{|l|l|}
\hline
\textbf{Parameter} & \textbf{Value} \\ \hline
Update rate main network &  5  \\ \hline
Update rate target network & 25 \\ \hline
Learning rate & 0.7 \\ \hline
Discount factor & 0.7 \\ \hline
Batch size & 10 \\ \hline
\end{tabular}
\end{table}

\begin{figure}
    \centering
    \includegraphics[width=0.48\textwidth]{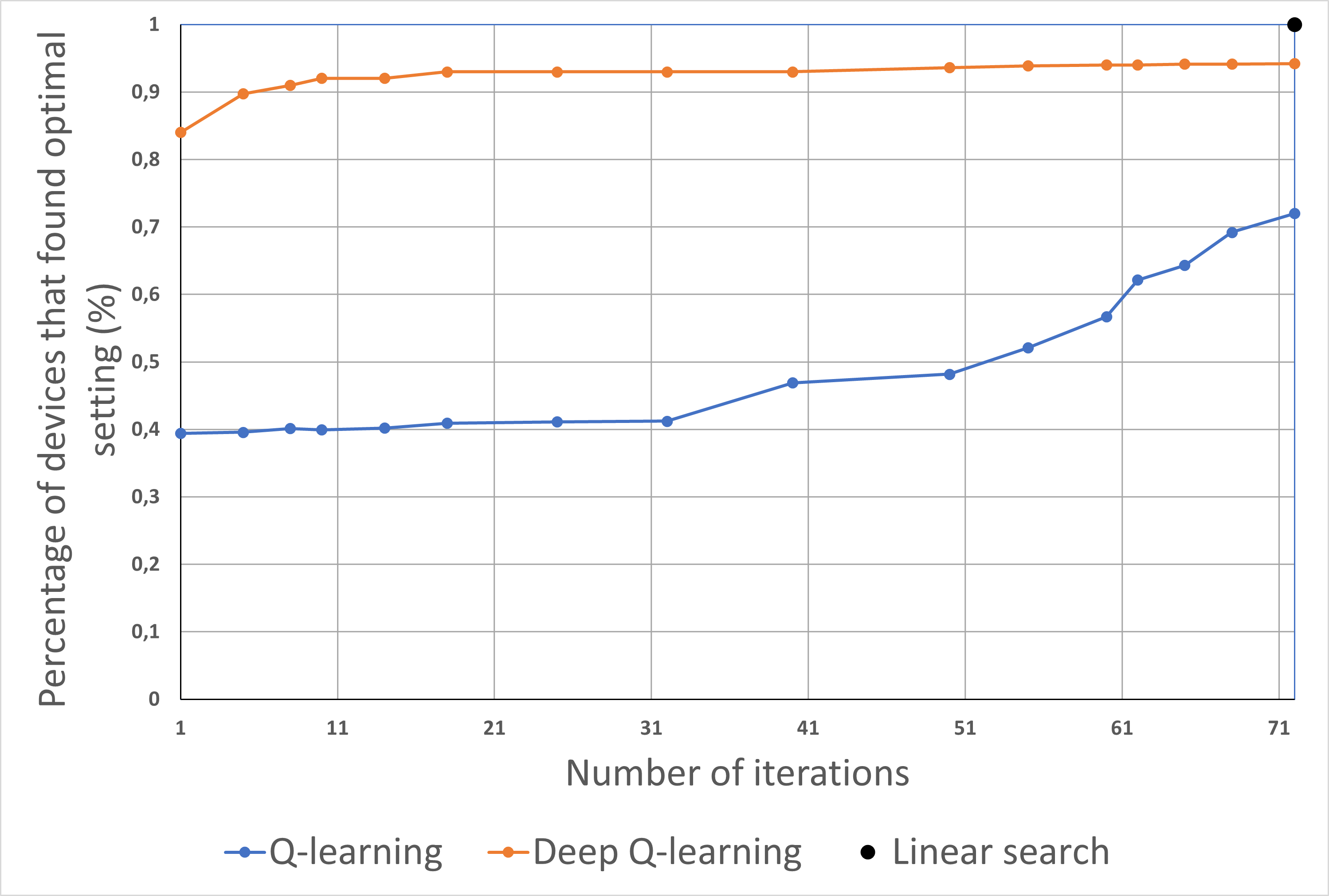}
    \caption{The percentage of optimal configurations selected in terms of number of iterations}
    \label{fig:staticcomp}
\end{figure}

In Figure \ref{fig:staticcomp}, the percentage of UWB links that found settings within  5\% of the highest possible reward (for that link) in function of the number of iterations is shown for Q-learning and Deep Q-learning.
The y-axis value is explained mathematically in (\ref{eq:opteq}), with $D$ the complete dataset, $l$ a link in between two UWB devices, $R$\_selected($l$) the reward of the UWB PHY setting selected by the algorithm for link $l$ and $R$\_best($l$) the reward of the setting with the highest reward for link $l$. A link is considered 'optimal' if the reward of the selected UWB PHY setting is within 5\% of the setting with the highest possible reward, $Optimal(l)$ returns 1 in this case.
\begin{equation}
\label{eq:opteq}
    \textrm{Optimal(l)} =\begin{cases} 
    1 &\text{R\_selected(l) } \geq \textrm{R\_best(l)} \times 0.95 \\
    0 & \text{otherwise}
\end{cases}
\end{equation}

\begin{equation}
    \text{y} = \frac{\sum\limits_{\forall l \in D} \text{Optimal(l)}}{\sum\limits_{\forall l \in D} 1 }
\end{equation}

The performance of a linear search algorithm is indicated on the figure as well. This algorithm tries out every possible setting and selects the best performing one in the end. In this scenario, there are 72 possible settings, which means that the algorithm needs 72 iterations but is 100\% sure that the best setting is selected. Q- and deep Q-learning try to predict the best setting based on the link state parameters. Figure \ref{fig:staticcomp} clearly shows that Deep Q-learning performs better than Q-learning. After the first iteration, which means starting at a random action (iteration 0)  measuring the link state and then configuring a setting, the Q-learning algorithm selects an optimal setting for 39\% of the links between UWB devices and Deep Q-learning for 84\%. After 10 iterations, the percentage has increased to 92\% for Deep Q-learning, while the Q-learning is still only around 40\%. Comparing the algorithms when they have used the same amount of iterations as linear search shows that Deep Q-learning finds the best setting in 95\% of the cases and Q-learning 70\%. These results show the importance of using Deep Q-learning instead of Q-learning. The drawbacks of using Q-learning as discussed in \ref{sec:qlearning} are clearly visible in these results.

Deep Q-learning shows a clear advantage over the linear search algorithm, as it can select an optimal setting with a high accuracy after only a few iterations. While the linear search algorithm requires 72 iterations.

\subsubsection{Dynamic environment}
To test the performance of the algorithms in a dynamic environment, a test scenario is defined as shown in Figure \ref{fig:9thfloor}. To simulate a walk around the office, every 5 seconds, one of the anchor-tag UWB nodes is switched to abruptly change the situation and test how well the algorithm can adapt. Harder links, containing more obstacles, require more energy. The color used for the tag-anchor link indicates the minimal amount of energy necessary to enable communication, as shown by the legend. The parameters used during this experiment are shown in Table \ref{tab:dynamicparams}.

\begin{table}[ht]
\centering
\caption{Parameters used in dynamic test scenario for runtime adaptation of UWB PHY settings}
\label{tab:dynamicparams}
\begin{tabular}{|l|l|}
\hline
\textbf{Parameter} & \textbf{Value} \\ \hline
Ranging update rate & 50 Hz \\ \hline
Update rate main neural network & 20 \\ \hline
Update rate target neural network & 100 \\ \hline
Mini batch size &  256\\ \hline
Learning rate & 0.55 \\ \hline
Discount factor & 0.55  \\ \hline
\end{tabular}
\end{table}

\begin{figure*}[ht]
    \centering
    \includegraphics[width=0.95\textwidth]{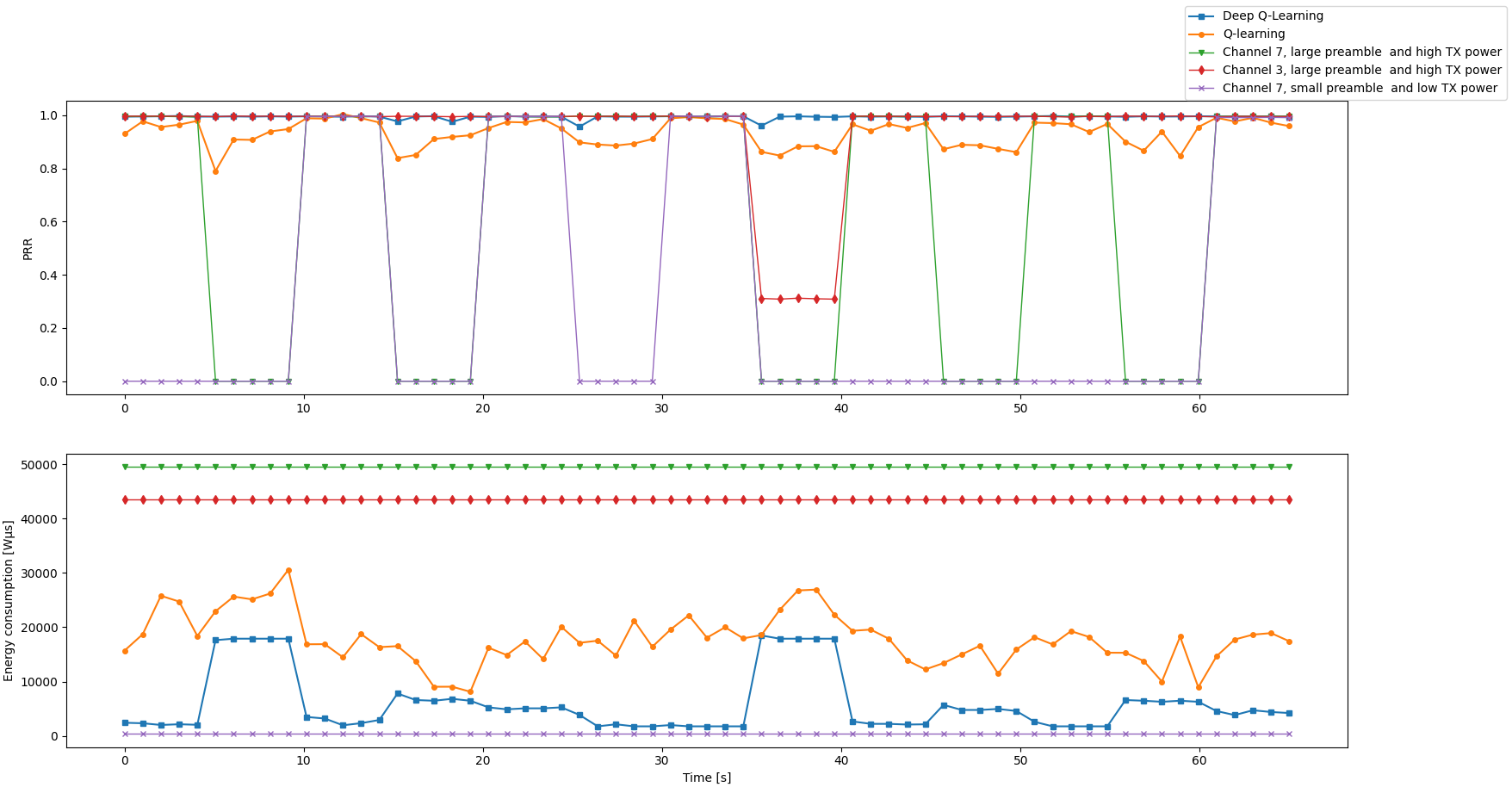}
    \caption{Comparison of the proposed Q-learning and deep Q-learning algorithms with constant high energy and low energy PHY settings in terms of PRR and energy consumption.}
    \label{fig:dynamicenvironment}
\end{figure*}

In Figure \ref{fig:dynamicenvironment}, the PRR and the energy consumption during this test scenario are shown for different cases, namely deep Q-learning, Q-learning and three fixed PHY settings. These were chosen because they represent the difference in performance of high and low energy settings and the influence of the channel.The fixed PHY settings are (1) a high energy consuming setting on channel 7 \{7, 4096, 64, 6800, 10.5\}, (2) a high energy consuming setting on channel 3 \{7, 4096, 64, 6800, 10.5\} and (3) a low energy consuming setting on channel 7 \{7, 128, 64, 6800, 0\}.
Figure \ref{fig:dynamicenvironment} shows that using a fixed PHY setting does not guarantee good reliability. Communication drops off completely at several links for the fixed PHY settings on channel 7, even for the high energy consuming one. The high energy consuming setting on channel 3 has better reliability, except for the single drop to 40\%.

Q-learning has no links where the communication drops off completely, but fails to provide true reliability as the PRR is never 100\%. This means that frames are continuously lost. However, it still uses considerably less energy than the two high energy consuming ones. However, it does not select very low energy consuming settings.

Figure \ref{fig:dynamicenvironment} shows that the Deep Q-learning algorithm can ensure reliable communication, by keeping the PRR constantly very close to 100\%, while dynamically selecting settings with higher energy consumption when necessary. This is well demonstrated by the sharp increase in energy consumption at time intervals 5-10 and 35-40 seconds of the evaluation. In these time intervals, the UWB communication is happening over a red link (number 2 and 8) in \ref{fig:9thfloor}. Between 40 and 65 seconds, we can clearly see that the algorithm can also make the distinction between links with a smaller difference in required energy consumption (green vs. yellow). However, deep Q-learning never selects a PHY setting with an energy consumption as low as the low energy setting, even though this setting can enable reliable communication for certain links. 

Table \ref{tab:avgperf} compares the average performance of the five different cases in the figure in terms of PRR, energy consumption and ranging error.
\begin{table*}[ht]
\centering
\caption{Average performance comparison of the different methods in the test scenario }
\label{tab:avgperf}
\begin{tabular}{cccc}
\textbf{Fixed radio configuration / Proposed algorithm} & \multicolumn{1}{c}{\textbf{Average energy}} & \multicolumn{1}{c}{\textbf{Average}} & \multicolumn{1}{c}{\textbf{Average}} \\
\textbf{} & \multicolumn{1}{c}{\textbf{consumption (W$\mu$ s)}} & \multicolumn{1}{c}{\textbf{PRR (\%)}} & \multicolumn{1}{c}{\textbf{ranging error (mm)}} \\ \hline
\textbf{Channel 7, small preamble  and low $P_{tx}$} & 373.92 & 30.59 & 295.018\\ \hline
\textbf{Channel 7, large preamble and high $P_{tx}$} & 49572.15 & 61.23 & 318.02\\ \hline
\textbf{Channel 3, large preamble and high $P_{tx}$} & 43497.42 & 94.25 & 235.92\\ \hline
\textbf{Q-learning} & 17782.15 & 93.26 & 243.45\\ \hline
\textbf{Deep Q-learning} & 5898.92 & 99.31 & 240.65\\ \hline
\end{tabular}
\end{table*}
Deep Q-learning clearly performs best, it has the highest average PRR, the lowest energy consumption and good ranging error performance. The high energy consuming setting on channel 3 comes closest based on PRR and has slightly better ranging error performance, but it consumes more than seven times the amount of energy. While Q-learning performs considerably worse than deep Q-learning, it has a similar PRR and ranging performance as the high energy consuming setting on channel 3 and consumes less than half the amount of energy. The fixed settings on channel 7 perform the worst as the PRR is low, even when consuming a lot of energy. 

\subsection{Complexity analysis}
\subsubsection{Algorithmic complexity}
The complexity of the Q-learning and Deep Q-learning is mainly determined by the complexity of determining the Q-value approximation belonging to a certain state \cite{complex1} and updating the model or table. For Q-learning, this is simply selecting a row from a table $O(1)$ and changing one value in the table. For the Deep Q-learning, the complexity depends on the structure and hyperparameters of the neural network. For a neural network composed by fully-connected layers, the complexity is given by $O(mn$ log $n)$, with $m$ the number of layers and $n$ the number of units per layer \cite{complex2}.
\subsubsection{Time complexity}
For Q-learning, the time to calculate the optimal setting and update the Q-table based on the received information is dependent on the time to select the maximal value in a row and update one value in the Q-table. When executing the algorithm on a Windows laptop containing an Intel(R) Xeon(R) E3-1200/1500 v5/6th Gen, this takes on average 6.89 ms. For the Deep Q-learning algorithm, this time is dependent on the inference time of the neural network and the time to update the neural network with a new batch of data. This time depends on the size of the model, as well as the batch size and update rate of the neural network. The main factor contributing to this is the update rate of the model. As retraining, the model is the most extensive task. A higher update rate makes the model more dynamic, but has the drawback of longer time complexity. For the settings used during the evaluation shown in Table \ref{tab:dynamicparams} the average time is 23.18 ms, while the training time with one batch takes on average 320 ms.

\section{Conclusion}
\label{sec:conclusion}
For the increasingly prominent indoor localization technology UWB, accurate ranging has been extensively researched. However, enabling reliable, low energy consuming UWB communication in dynamic environments is largely unexplored. This work proposes a deep Q-learning algorithm for improved reliability in dynamic environments while minimizing energy consumption by changing the PHY layer settings based on link state measurements. This method outperforms using a fixed PHY layer setting and an exhaustive search, which are currently the most common ways to set the PHY setting. We found that deep Q-learning can achieve higher PRR while using considerably less energy and whilst reducing the ranging error. We also concluded that using traditional Q-learning does not suffice to solve this problem. This work can be extended by improving upon the current reward function to enable deep Q-learning to distinguish better between settings that have smaller differences in energy consumption. In addition, future work could employ RL for UWB localization systems using Time Difference of Arrival (TDoA) instead of two-way ranging or multi-agent localization problems. 
\section{Acknowledgment}
This work was supported by the imec ICON WISH project HBC.2021.0664.
\bibliographystyle{IEEEtran}
\bibliography{sample-base}
\newpage

\end{document}